\documentstyle[pre,aps,floats,epsf]{revtex}

\begin{document}

\renewcommand{\thefootnote}{\arabic{footnote}}
\draft

\twocolumn[\hsize\textwidth\columnwidth\hsize\csname@twocolumnfalse%
\endcsname

\title{Disorder effects in cellular automata for two-lane traffic}

\author{Wolfgang Knospe$^{1}$, Ludger Santen$^{2}$, Andreas
  Schadschneider$^{2}$, Michael Schreckenberg$^{1}$ \\}

\address{$^{1}$Theoretische Physik FB 10,
  Gerhard-Mercator-Universit\"at Duisburg, D-47048 Duisburg, Germany}
\address{$^{2}$ Institut f\"ur Theoretische Physik, Universit\"at zu K\"oln 
  D-50937 K\"oln, Germany}

\date{\today}

\maketitle              

\begin{abstract}
For single-lane traffic models it is well known that particle disorder
leads to platoon formation at low densities. Here we discuss the effect
of slow cars in two-lane systems. Surprisingly, even a small number of
slow cars can initiate the formation of platoons at low densities. 
The robustness of this phenomenon is investigated for different variants 
of the lane-changing rules as well as  for different variants on 
the single-lane dynamics. It is shown that anticipation of drivers 
reduces the influence of  slow cars drastically. 
\end{abstract}

]

\section{Introduction}
\label{Sec_Int}

In recent years it has turned out that cellular automata (CA) are excellent 
tools for the simulation of large scale traffic networks. The most prominent
example for this kind of models has been introduced by Nagel and
Schreckenberg~\cite{Nagel92} (NaSch model). Its properties have been 
discussed in detail in the past few years 
\cite{Schreckenberg95,Wolf95TGF,Schreckenberg97TGF}. 
Also more sophisticated CA have been developed, e.g. models with so 
called 'slow to start' rules, which are able to reproduce hysteresis 
effects in traffic flow \cite{Barlovic98}.

While most investigations consider homogeneous systems with one type of
cars on a translational invariant lattice, real traffic is
in many respects inhomogeneous. In general, different types of cars are
present and from daily experience it is known that slow cars have strong
influence on the systems performance. This is obviously most pronounced in 
single-lane traffic, where passing is not possible and therefore the slow 
cars dominate the dynamics. This intuitive picture has been confirmed
analytically \cite{Krug96,Krug97,Evans96,Evans} for the asymmetric
exclusion process, which is
closely related to the NaSch model with
$v_{max}=1$, and numerically~\cite{Ktitarev97} for the NaSch model with
$v_{max} >1$.
Beyond the basic mechanism of platoon formation behind the slowest
car, it has been shown that a phase transition occurs, which is in
some sense similar  to the Bose-Einstein condensation \cite{Evans96}. 
For multi-lane models the influence of slow cars is not so
obvious. Recently is has been shown, that already a small fraction of
slow cars can dominate multi-lane systems at low
densities~\cite{Chowdhury97}, but nevertheless the effect of
different types of cars in multi-lane traffic is far from being clarified.

In this work we show that for very small densities of slow cars
platoon formation is observable, if one considers the  CA
model for two-lane traffic introduced by Rickert et al.~\cite{Rickert2}. 
Moreover we discuss
alternative lane-changing rules as well as variants of the basic model.

For the sake of completeness we briefly recall the definition of the
NaSch model \cite{Nagel92}. The NaSch model is a discrete model for traffic 
flow. The road is divided into cells which can be either empty or occupied 
by a car with a velocity $v=0,1,...,v_{max}$. The cars move from
the left to the right on a lane with periodic boundary conditions and the
system update is performed in parallel 
for all cars according to the following four rules:
\begin{enumerate}
\item Acceleration: $v \rightarrow \min(v+1,v_{max})$.
\item Deceleration: $v \rightarrow \min(v,gap)$
\item Noise: $v \rightarrow \max(v-1,0)$ with probability $p$.
\item Motion: $x \rightarrow x+v$.
\end{enumerate}
$v$ denotes the velocity, $v_{max}$ the maximum velocity and $x$ the
position of a car, $gap$ specifies the number of empty
cells in front of the car.

In order to extend the model to multi-lane traffic one has to introduce
lane-changing rules. This is usually done by dividing the update step 
into two sub-steps:
In the first sub-step, cars may change lanes in parallel 
according to lane-changing rules and in the second sub-step the lanes
are considered as independent single-lane NaSch models.

The lane-changing rules can be symmetric or asymmetric with respect to
the lanes and to the cars. Rickert et al.~\cite{Rickert2} have assumed a
symmetric rule set where cars change lanes if the following two criteria
are fulfilled:
\begin{itemize}
\item   Incentive criterion:
\begin{enumerate}
\item $v_{hope} > gap$,\ \ with $v_{hope} = \min(v+1,v_{max})$.
\end{enumerate}
\item Safety criteria:
\begin{enumerate}
\item [2.] $gap_{other} > gap$.
\item [3.] $gap_{back} \ge v_{max}$.
\end{enumerate}
\end{itemize}
Here $gap$, $gap_{other}$, $gap_{back}$ denote the number of free cells
between the car and its predecessor on the actual lane and its
two neighbor cars on the desired lane, respectively.

Recently, asymmetric rule sets have been proposed for the description
of highway traffic on the german ``Autobahn`` where overtaking on the
right lane is forbidden. These rule sets are able to reproduce the 
lane-usage inversion\footnote{Above a certain density most of the cars are on
the left lane.} observed experimentally in excellent agreement 
with measurements
\cite{Rickert2,nagelzweispur,Wagner97,Wagnerzwei,Nagatani96,Nagatani96a,Chowdhury97}.

The outline of this paper is as follows:
In the next section the lane change behavior of a homogeneous NaSch
model is considered. It will be shown that the lane-change probability
decreases for certain braking parameters $p$. This leads to a total
domination of the system not only by a fraction of slow cars, but also
by just one slow car. Therefore we will first change the velocity update
order and consider a sequential version of the NaSch model in section 
\ref{Sec_ModMo}. We also present a parallel ``anticipation`` model which 
shows features found in the basic parallel as well as in the sequential 
model. Finally, several other variants of the basic model are discussed
briefly. In the last section a short summary and a discussion will follow.


\section{Basic models}
\label{Sec_models}

As mentioned above it is necessary to choose an appropriate lane-changing 
rule set depending on the experimental situation. First we consider
symmetric lane-changing rules, which are relevant for traffic in towns
and on highways, where overtaking on both lanes is allowed.

\begin{figure}[hbt]
\epsfxsize=\columnwidth\epsfbox{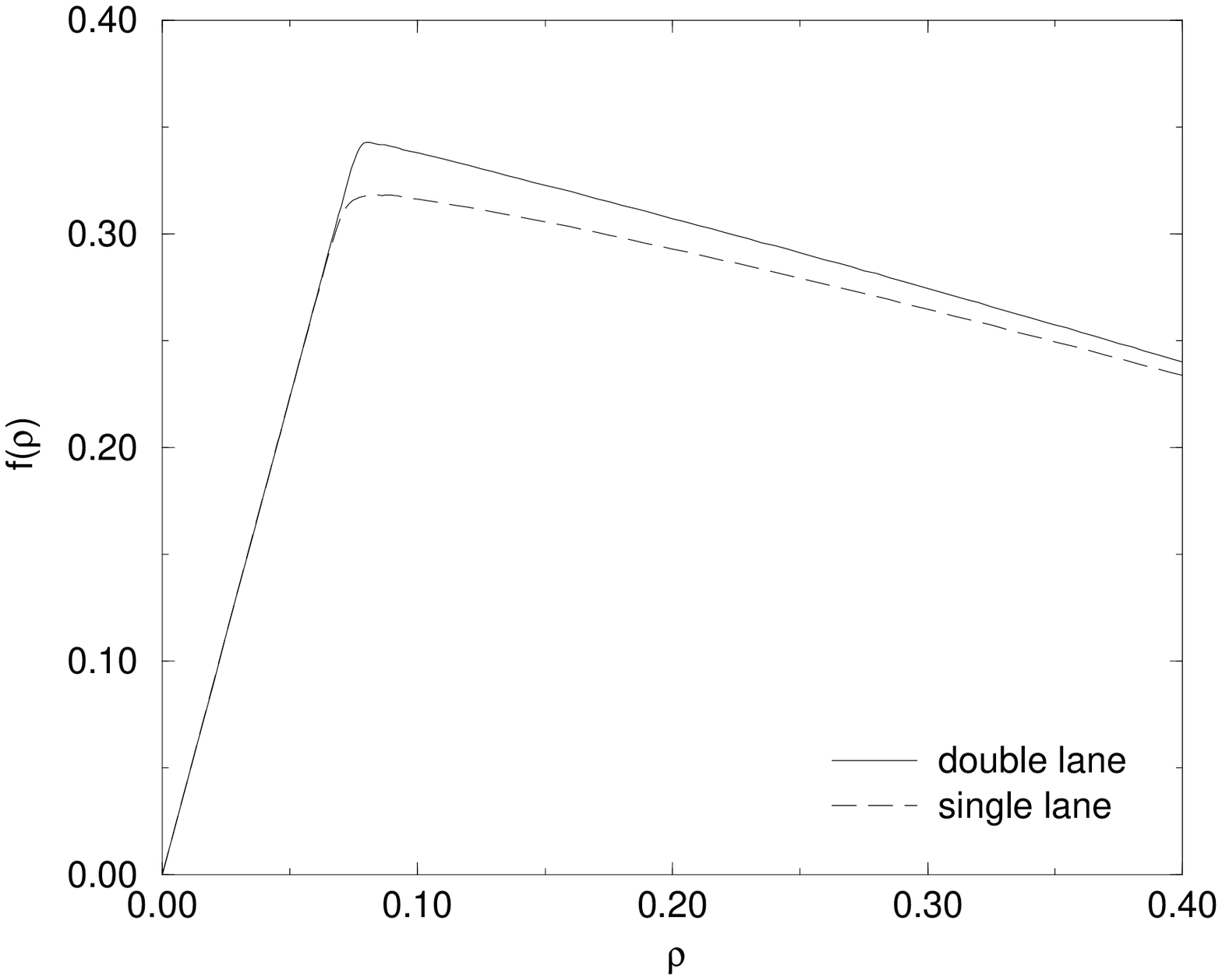}
\epsfxsize=\columnwidth\epsfbox{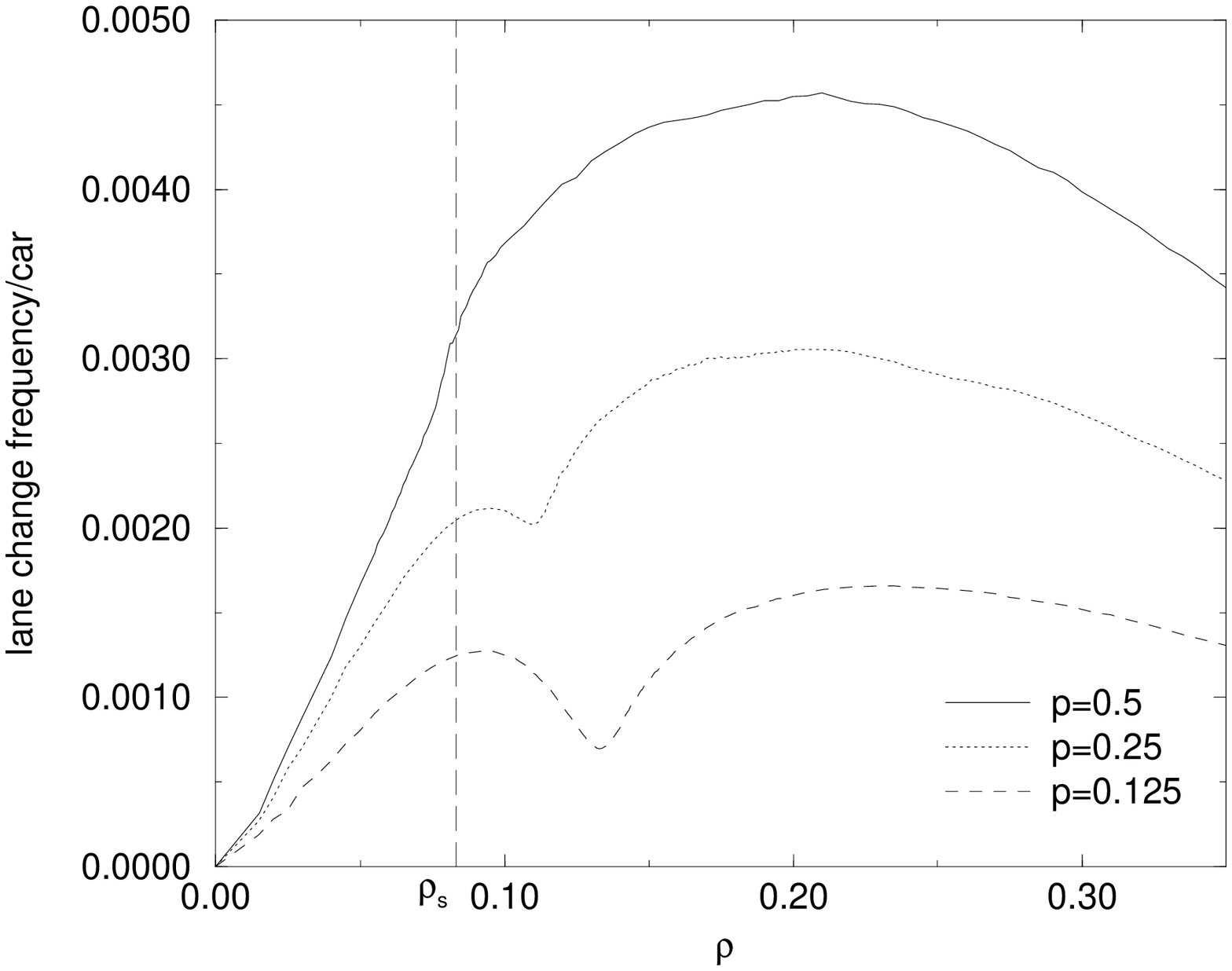}
\caption{Top: Flow per lane of the single lane model
compared with the two lane model for systems with $v_{max} = 5$ and $p
= 0.5$. Bottom: Lane change frequency in the two lane model for
different braking parameters $p$.}
\label{abb1}
\end{figure}

Fig.~\ref{abb1} shows the fundamental diagram of a periodic two-lane system
and the density dependence of the lane-changing frequency. The simulations 
reproduce well known results, e.g.\ an increase of the maximum flow per 
lane compared to the flow of a single-lane road. In addition, one finds
suprising new results like a local minimum of the lane-changing frequency 
near the density of maximum flow for small braking probabilities $p$. 
Obviously jams can be partially avoided by changing the lanes 
and therefore the capacity of the system is slightly increased.

\begin{figure}[htb]
\epsfxsize=\columnwidth\epsfbox{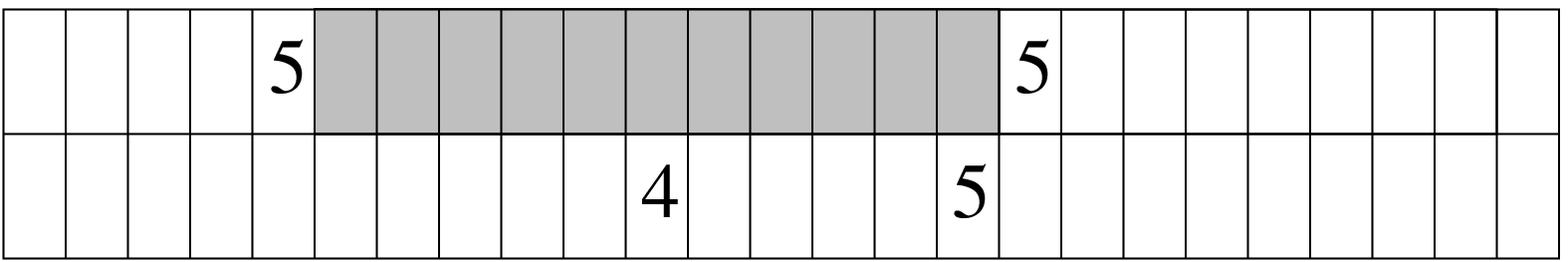}
\epsfxsize=\columnwidth\epsfbox{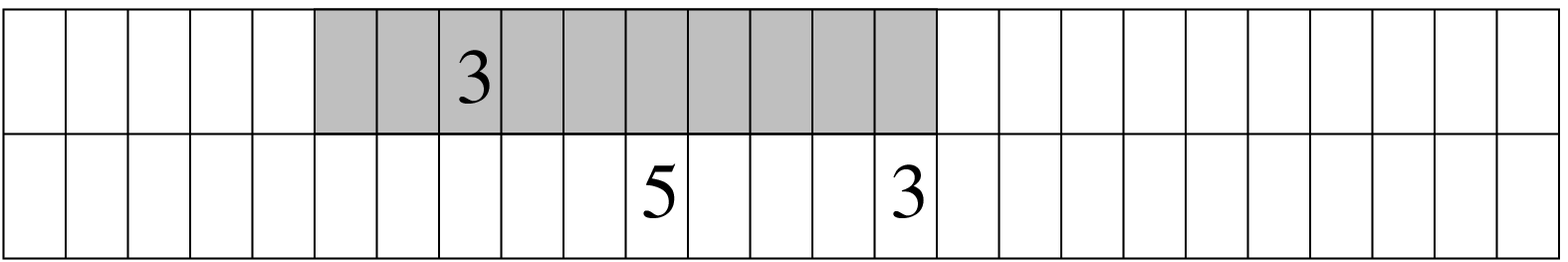}
\caption{Top: Necessary safety gap for a lane change in the free
flow regime.
Bottom: Possible positions of slow cars which lead to a
plug. The cars are driving from left to right.}
\label{abb3}
\end{figure}

The behavior of the lane-changing frequency can be explained if one
takes into account the number of empty cells necessary for a
lane-changing procedure.
Two prerequisites have to be fulfilled in order to initiate a lane
change. First, the situation on the other lane must be more convenient
and second, the safety rules must be fulfilled. Therefore
one needs typically  $2v_{max}+1$ empty cells on the destination
lane for a lane-changing maneuver in the free flow regime (Fig.~\ref{abb3}).
Hence, one finds a local maximum of the lane-changing frequency near
$\rho_{s} = \frac{1}{2} \frac{1}{v_{max}+1}$ if the cars are
ordered homogeneously, which typically happens for small values of $p$.
For larger values of the braking noise  (i.e. $p=0.5$) no local
maximum is observable.
Increasing the density for sufficiently small values of $p$, one finds a
pronounced minimum of the lane changing frequency. This can be understood
in the limit $p\to 0$ where, for $\rho = \frac{1}{v_{max}+1}$, the cars are
perfectly ordered with a gap of $v_{max}$ sites between consecutive vehicles.
Obviously in this case both the incentive and the safety criteria are never
fulfilled and the lanes are completely decoupled. For small noise the
ordering mechanism is still present and therefore the number of lane changes
is drastically reduced near $\rho = \frac{1}{v_{max}+1}$. For large values
of $p$ this kind of ordering is suppressed since fluctuations of the distance
between consecutive cars become larger. Therefore large gaps become more
likely than in the deterministic limit and lane changes become possible 
again. A further increase of
the average density reduces the probability to find a gap on the
other lane which is large enough for a lane change. Obviously the global 
maximum is reached at much higher values
of $\rho$, because the incentive criterion is fulfilled for
more and more cars at higher densities, and at intermediate densities
large gaps are still present because the microscopic states are
inhomogeneous.

\begin{figure}[hbt]
\epsfxsize=\columnwidth\epsfbox{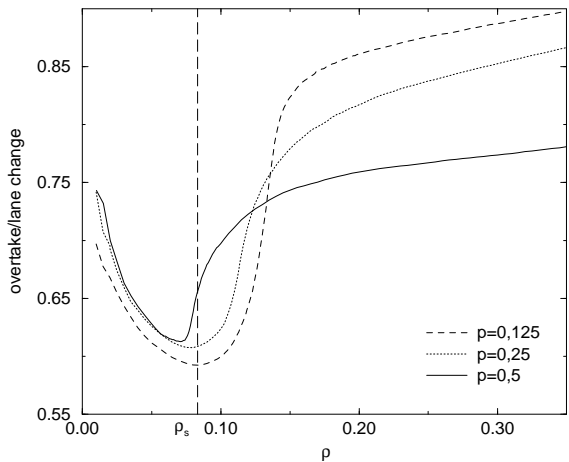}
\caption{Fraction of lane changes which are overtaking maneuvers for
different braking parameters $p$.}
\label{abb4}
\end{figure}

This intuitive picture can be confirmed by measuring the number of
``successful`` lane changes, where a car actually overtakes its former
predecessor (Fig.~\ref{abb4}) in the next timestep. Overtaking means that
following and leading car have change their r\^ole. Again we find 
a local minimum near $\rho_{s}$. For  
densities $\rho > \rho_{s}$ the efficiency of lane changes increases
monotonously, meaning that the cars can improve their average velocity
due to changing the lane.

After this short review of the results for homogeneous systems, we now
consider different types of cars which is obviously more relevant for
practical purposes. As a first step towards realistic distributions of
free flow velocities we have chosen two types of cars, e.g.\ slow cars
(`trucks') with $v_{max}^{slow}=3$ and fast cars with
$v_{max}^{fast}=5$, analogous to Ref.\cite{Chowdhury97}. 
The simulations have been carried out with $5\%$ of slow cars, which are
initially positioned randomly. The fast as well as the slow cars may use 
both lanes, e.g.\ both cars are treated equally with respect to the 
lane-changing behavior.

\begin{figure}[ht!]
\epsfxsize=\columnwidth\epsfbox{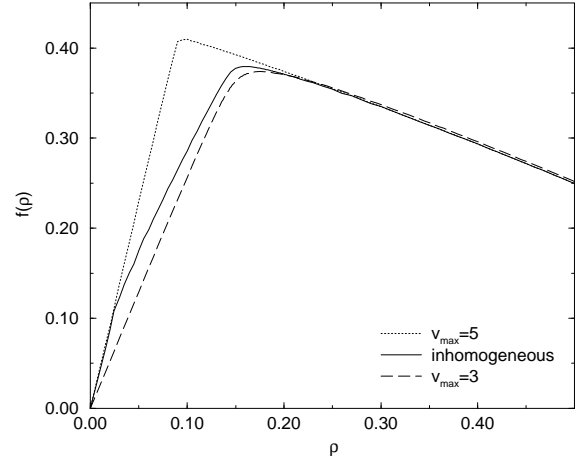}
\caption{Comparison of the flow per lane of the inhomogeneous model with 
the corresponding homogeneous models for $p=0.4$.}
\label{abb4a}
\end{figure}

In Fig.~\ref{abb4a} the effects of the slow cars on the average flow of 
the two-lane system is compared with the fundamental diagram of a single
lane road with one slow car. Since passing is not allowed for 
a single-lane system, clearly the slow car dominates the average flow 
at low densities and platoon formation is observable
\cite{Krug96,Krug97,Evans96,Evans,Ktitarev97}. Surprisingly the two-lane 
system shows a quite similar behavior, although passing is
allowed and the fraction of slow cars is rather small, which is
consistent with the results of Ref.\cite{Chowdhury97}.

In~\cite{Rickert2} it was shown, that two trucks driving side by side
can form a ``plug`` and blockade the following traffic. Similar observations
have been made in \cite{Chowdhury97,nagelzweispur,Huberman}.
For constituting such a plug it is not necessary that both cars are
driving side by side. A fast car with velocity $v=v_{max}^{slow}$
driving behind a slow car needs $v_{max}^{slow}+1$ empty cells in the
driving direction and $v_{max}^{fast}$ empty cells as safety gap on
the other lane for a successful lane change. Therefore two trucks are
able to form a plug even with a gap of $9$ cells between them 
(Fig.~\ref{abb3}).

These plugs lead to platoon formation analogous to a single-lane
system. Obviously the average velocity of such platoons is limited by
the free flow velocity of the slow cars. Therefore, if the plugs have 
long lifetimes, the total flow can not exceed the capacity of a homogeneous 
system of slow cars. 

\begin{figure}[!ht]
\epsfxsize=\columnwidth\epsfbox{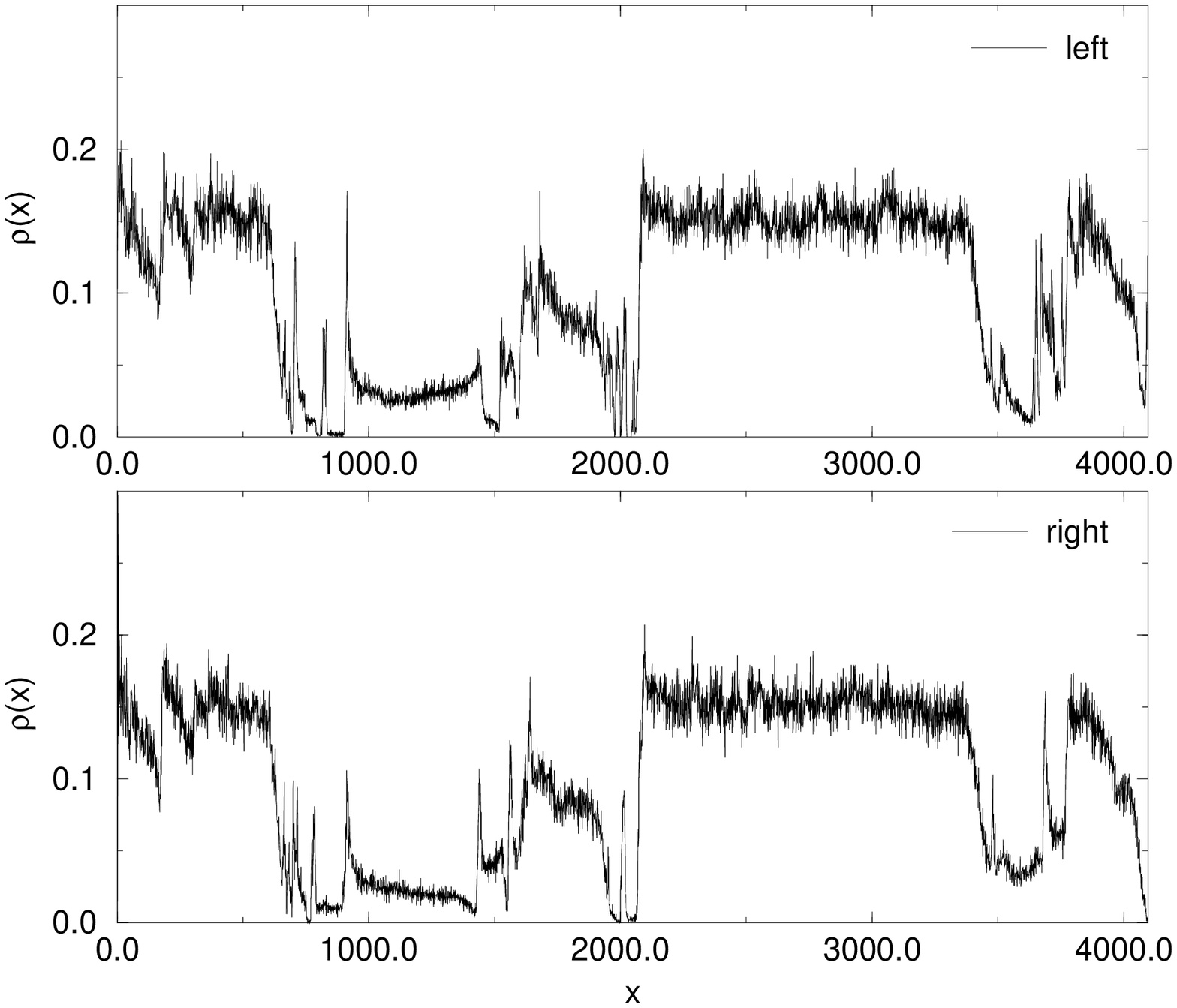}
\epsfxsize=\columnwidth\epsfbox{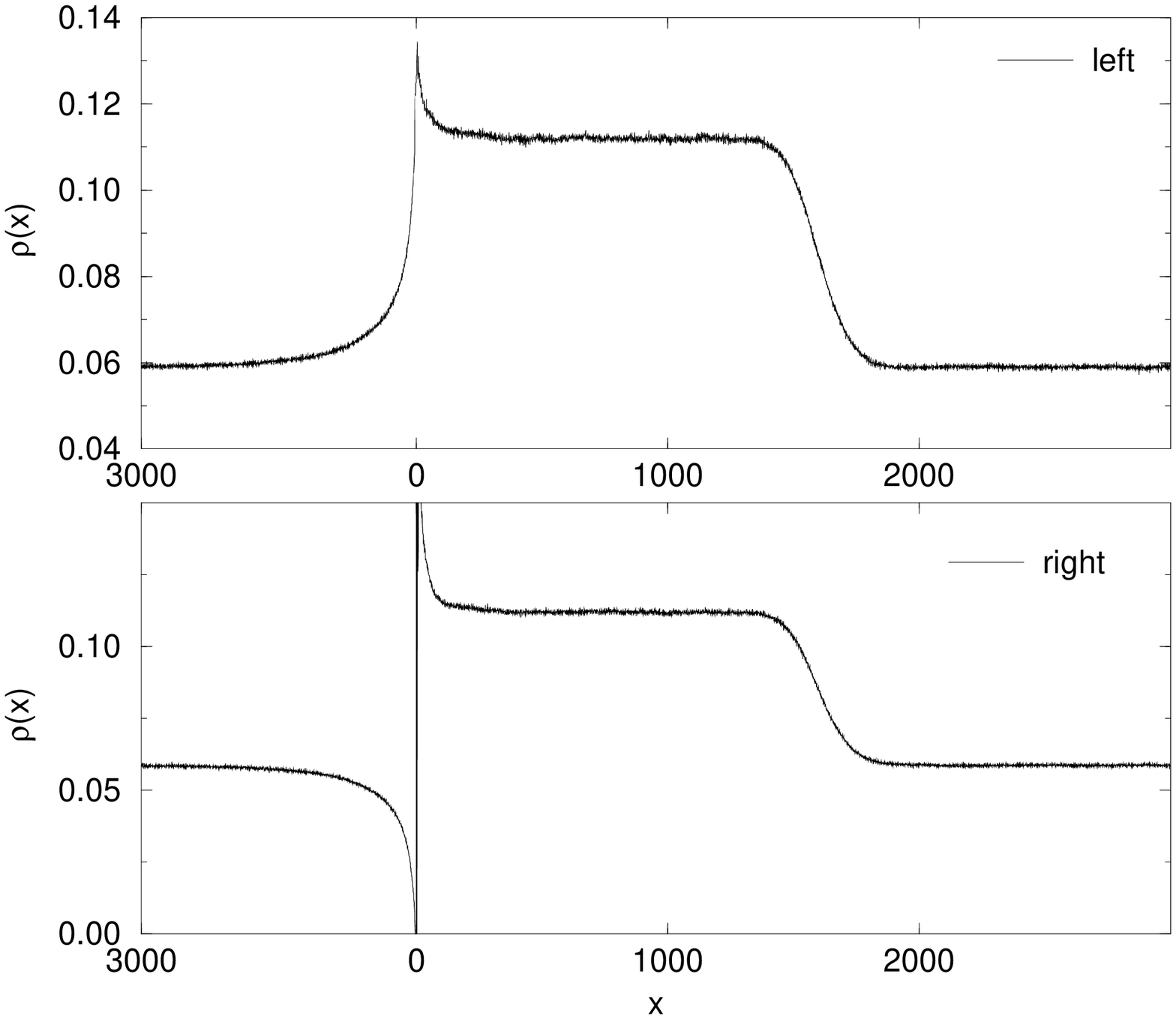}
\caption{Top: Density profile relative to a slow car on one lane in the
inhomogeneous model (with $5\%$ slow cars) at $\rho = 0.1$ and $p=0.4$.
Bottom: Density profile of both lanes in a system with just one slow
car relative to the slow car for $\rho=0.08$ and $p=0.125$. The system 
size in both cases is $L=4096$.}
\label{abb7}
\end{figure}

In fact, plug configurations are quite stable because the slow cars 
generically will have a large gap in front such that the incentive 
criterion hardly ever will be fulfilled. Therefore the leading vehicle 
of a platoon rarely changes the lane. Moreover both leading cars of the 
plug drive with the same average velocity and therefore larger distances 
between the slow cars only occur due to velocity fluctuations. This means 
that plugs have strong influence on the stationary state. 

In many respects the situation is analogous to single-lane traffic. This
can be exemplified by looking at the density profiles of both 
lanes relative to a slow car on one of the lanes. Obviously the flow is
dominated by plugs. Beneath several small jams one big jam parallel
on both lanes occurs, comparable to the density profile of a 
single-lane road (Fig.~\ref{abb7}).
Note, however, that in contrast to one-lane systems \cite{Krug96}
the variance of the distance distribution does not diverge.

For densities $\rho > \rho_{max}$, where $\rho_{max}$ is the density
where the flow becomes maximal, also the slow cars will not always 
find a sufficient large gap in front. Therefore the average flow is limited 
by the number of empty cells but not by the low maximum velocity of the slow
cars. Consequently we get the same flow for the homogeneous system of
fast as well as for the inhomogeneous case.

\begin{figure}[!ht]
\epsfxsize=\columnwidth\epsfbox{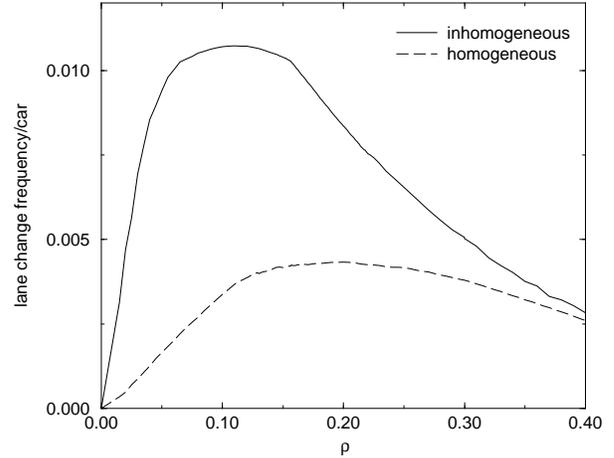}
\epsfxsize=\columnwidth\epsfbox{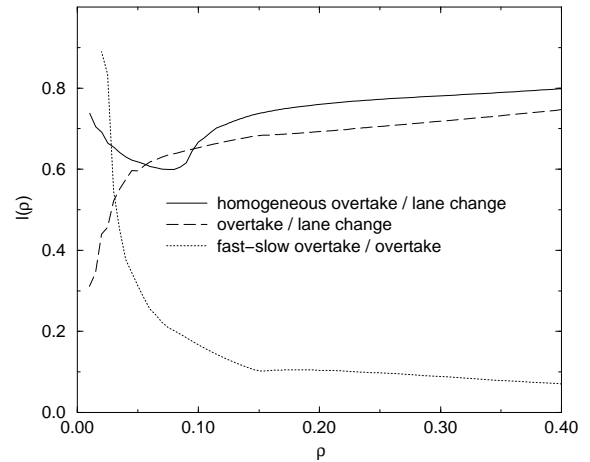}
\caption{Top: Lane-changing frequency of the inhomogeneous model compared to
the corresponding homogeneous model for $p=0.4$.
Bottom: Fraction $l(\rho)$ of lane changes which are overtaking maneuvers
in the homogeneous (full line) and inhomogeneous system (broken) line.
The dotted line shows the fraction of overtaking maneuvers where a slow
car is passed by a fast car.
$p=0.4$ in all cases.}
\label{abb8}
\end{figure}


Fig.~\ref{abb8} shows a comparison of the lane-changing frequency and
the fraction of successful overtaking maneuvers for the homogeneous
and the inhomogeneous models. For small densities the lane-changing
frequency in the inhomogeneous model is increased drastically. For
very small densities most of the overtaking maneuvers are those
where fast cars pass a slow car. This corresponds to the regime
where the trucks do not have a significant effect on the flow.
With increasing density the fraction of fast-slow overtaking
maneuvers decreases although the fraction of successful overtaking
maneuvers increases.

As a next step we study the effect of a {\em single} slow car on the behavior
of the system. This slow car with $v_{max}^{slow}=3$ always moves on the
same (right) lane, i.e.\ it is not allowed to change the lane.
In contrast to the former example, one would expect that this slow car
disturbs the system only locally and therefore should not have any
effects on e.g.\ the fundamental diagram.

\begin{figure}[!ht]
\epsfxsize=\columnwidth\epsfbox{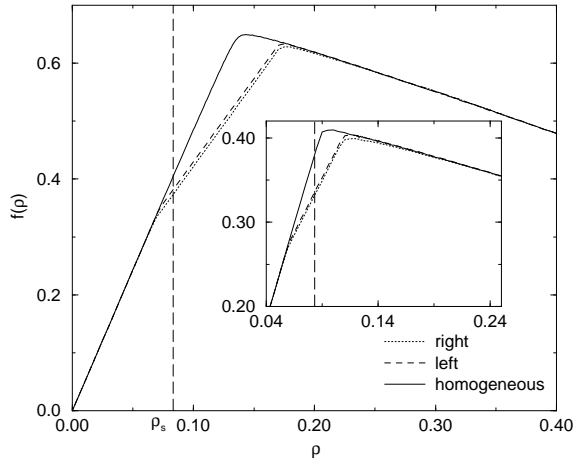}
\caption{Comparison of the flow per lane of a two-lane system
with one slow car with the homogeneous system, $p=0.125$. The inset
shows the same situation for $p=0.4$.} 
\label{abb10}
\end{figure}

\begin{figure}[hbt]
\epsfxsize=\columnwidth\epsfbox{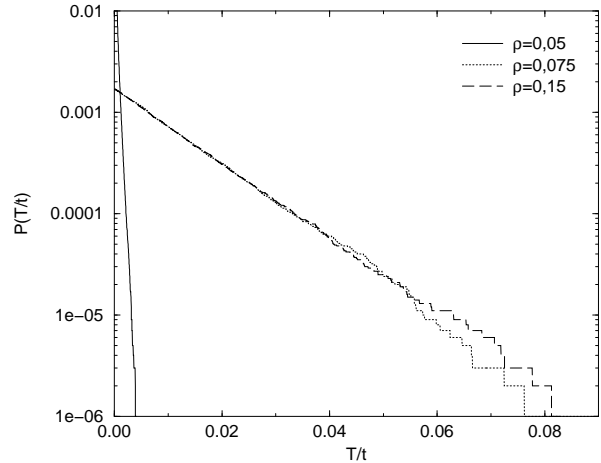}
\caption{Waiting time $T$ distribution for $p=0.125$. $t$ denotes the 
measure time.} 
\label{abb10a}
\end{figure}

Simulations show, however, that above a certain density $\rho_{T}$
already one slow car is sufficient to dominate the flow on both 
lanes (Fig.~\ref{abb10}). The transition density $\rho_{T}$ depends 
on the maximum velocity of the fast cars and the braking parameter $p$. 
While $\rho < \rho_{T}$ the fast cars change to the left 
lane in order to avoid being trapped behind the slow car. 
With increasing density the probability for a lane change decreases
and a jam behind the truck forms.
This jam causes a jam on the left lane (Fig.~\ref{abb8}). 
More and more cars change to the left and form a region of high
density parallel to the jam.
This region is disturbed if a jammed fast car can change to the left.
The local defect, the slow car, causes a ``parallel`` jam on the
left lane. 
Together with the truck this parallel jam behaves like a plug.
Therefore the flow on both lanes is dominated by this plug for
$\rho > \rho_{T}$, the slow car ``synchronizes`` the flow on both lanes.
The effect of vehicles trapped behind the slow car can be illustrated
by measurements of the waiting time distribution of the first fast car 
behind the slow one until it can change the lane (Fig.~\ref{abb10a}).
Close to $\rho_{T}$ the waiting time $T$ jumps to macroscopic values.

The choice of a low braking parameter $p$, on the one hand, leads to
a decrease of the lane change probability and therefore to an
influence even on the homogeneous system. 
On the other hand, with increasing $p$ interactions between the cars
increase. A slow car can easily cause a jam and form a
``plug``. Hence, the system is dominated by one
slow car above a certain density $\rho_{T}$ for a wide range of
braking parameters $p$.
With decreasing $p$ one observes $\rho_{T} \rightarrow \rho_{s} \approx
\frac{1}{2}\frac{1}{v_{max}+1}$. 

\begin{figure}[hbt]
\epsfxsize=\columnwidth\epsfbox{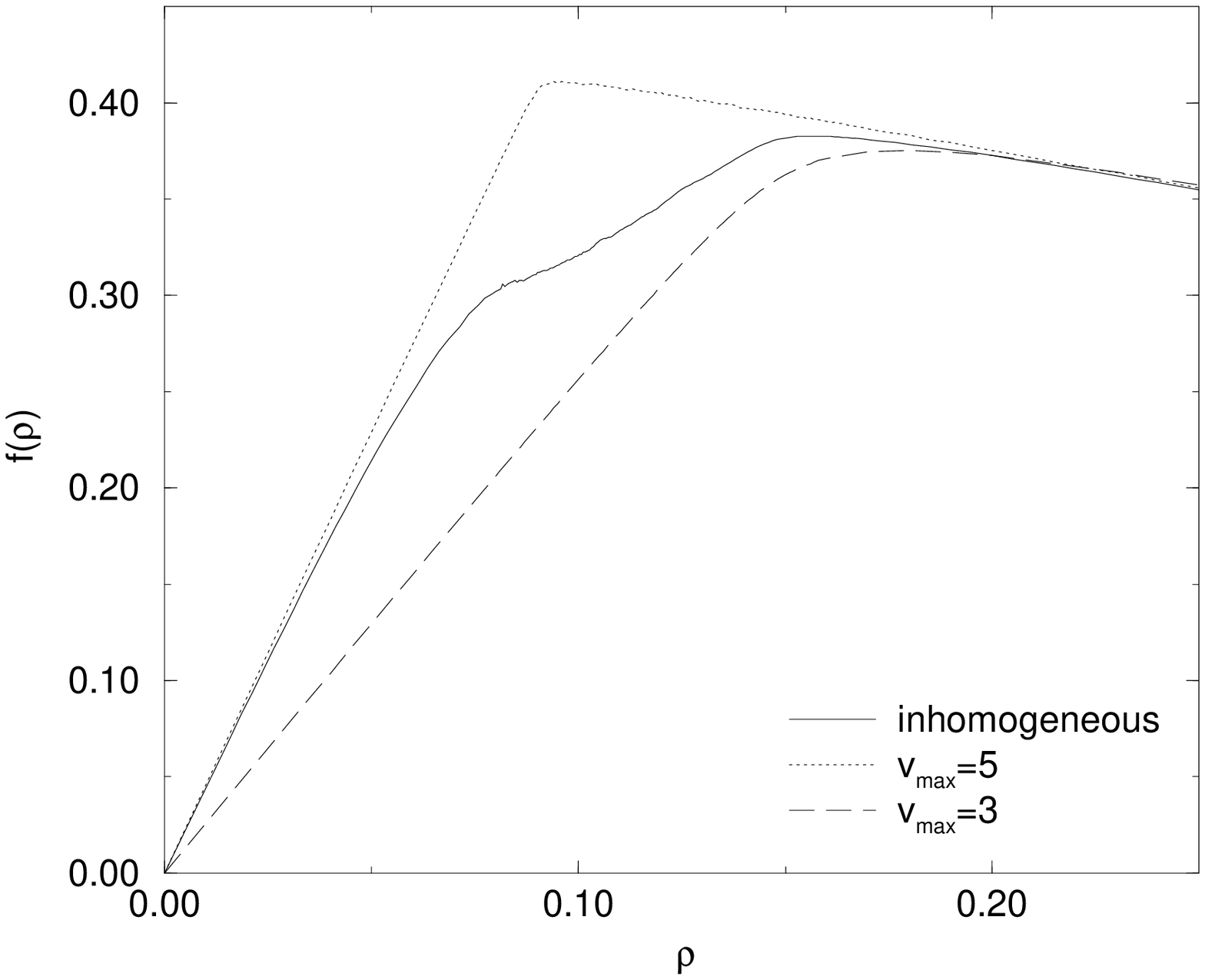}
\epsfxsize=\columnwidth\epsfbox{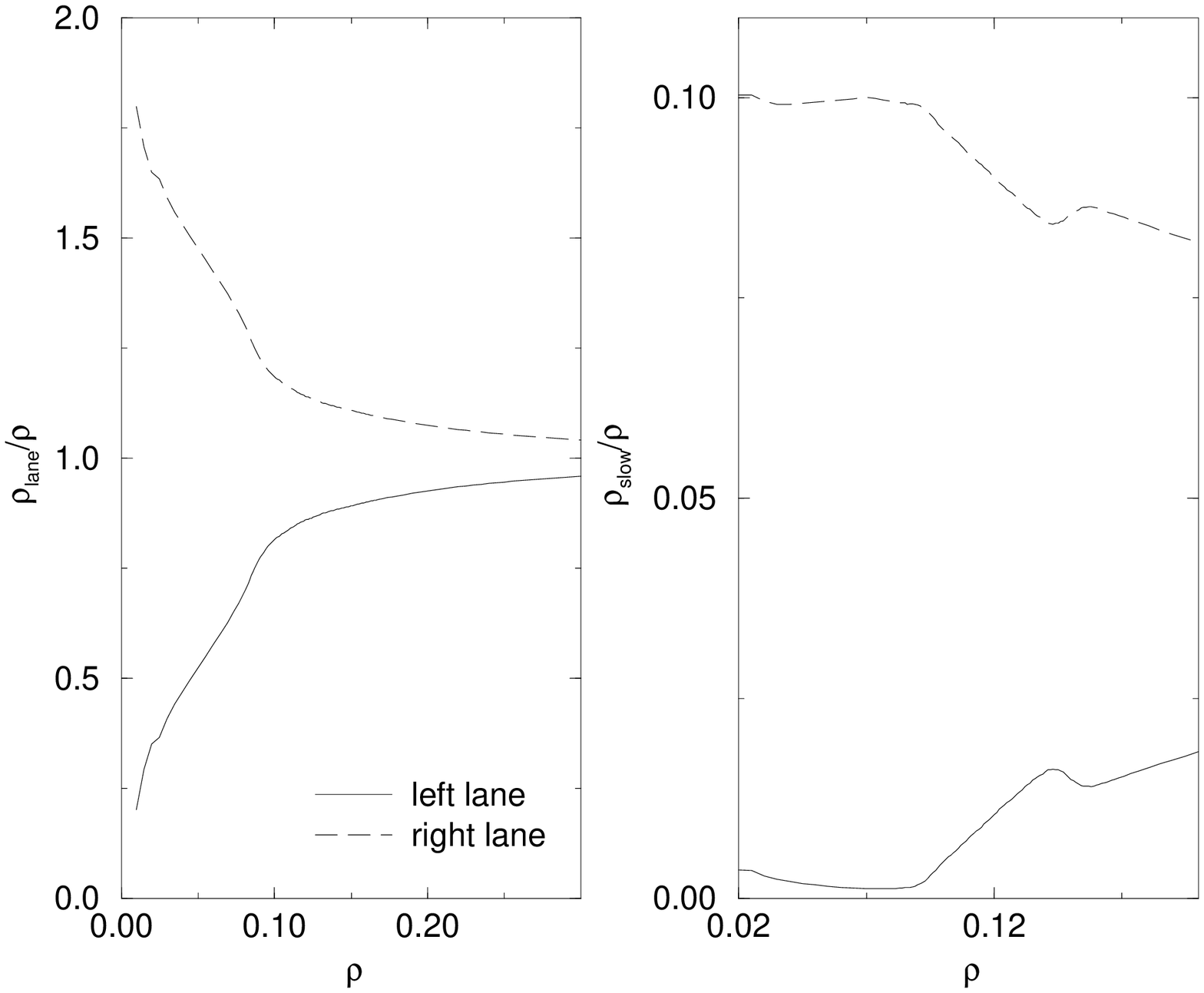}
\caption{Top: Average flow of different asymmetric models with
  $p=0.4$.
  Bottom: Relative densities of the lanes and slow cars for a
  inhomogeneous model with $5\%$ slow cars.}
\label{asym1}
\end{figure}

In order to weaken the influence of the slow cars one can introduce asymmetric
lane-changing rules. This can be done by neglecting the first rule of
the symmetric rule set of Rickert et al.~\cite{Rickert2} for the change 
from the left to the right lane.
Now, the left lane is designated as the passing lane and cars are
trying to change back to the right lane as soon as possible.
In Fig.~\ref{asym1} the average flow of asymmetric models with
$v_{max}=5$, $v_{max}=3$ and $v_{max}^{fast}=5$ with $5\%$ slow cars
($v_{max}^{slow}=3$) is compared.
For very low densities, fast cars can pass slow cars more effectively 
if asymmetric lane-changing rules are applied.
Obviously, the flow can be increased for densities below $\rho_{max}$ 
although the strong influence of the slow cars is still apparent.

With asymmetric rules almost all cars drive on the right lane
at low densities (Fig.~\ref{asym1}). 
Hence, the system is divided into a ``fast`` and a ``slow`` lane.  
Therefore at low densities there is only a low probability for plug
formation of two slow cars driving side by side. With increasing
density this probability and the probability for plugs formed by a
slow car and its parallel jam increases, such that platoon formation
occurs and the system is dominated by slow cars. Note that the density
on the fast lane is still reduced in that regime, in contrast to
realistic systems, where lane inversion is observable. Therefore the
increase of the system's performance is in some sense artificial. 

The results presented up to now show that even very small densities
of slow cars can dominate the behaviour of the whole system. Since
this appears to be somewhat unrealistic we will investigate in the
following several modifications of the basic rule set where the
influence of slow cars is reduced.

\section{Modified models}
\label{Sec_ModMo}

\subsection{Sequential update}

The results of the previous section have shown, that the safety
criteria demand large gaps between consecutive cars on the destination
lane for lane-changing maneuvers. This leads to plug formation and
therefore a drastic flow reduction by slow cars already in the
free flow regime. Much smaller gaps would be sufficient for a lane
change, if the driver takes into account the behavior of the
predeccesor in the next time step. Such anticipation effects can be
most easily considered in a discrete model, if the update procedure is
performed sequentially against the driving direction. This means that
the driver has full information about the behavior of the
predecessor in the next time step.

In order to estimate the adequacy of this approach we briefly discuss
the single-lane properties of the sequential variant of the NaSch
model. Using sequential update one obviously cannot ensure translational
invariance. E.g. if one performs the update site by site in a fixed
sequence, one generates effectively a bottleneck situation analogous
to a system with defect sites. Due to this shortcoming of a site oriented
implementation of the update sequence, we implemented the update
procedure in the following manner. The lattice update of one time step
starts at a "NaSch-car", i.e. a car with $v=0$ or $gap>v$. In the
following the position of all cars are updated car by car against the
driving direction\footnote{Using a fixed sequence of cars changes the
results only slightly.}.

\begin{figure}[hbt]
\epsfxsize=\columnwidth\epsfbox{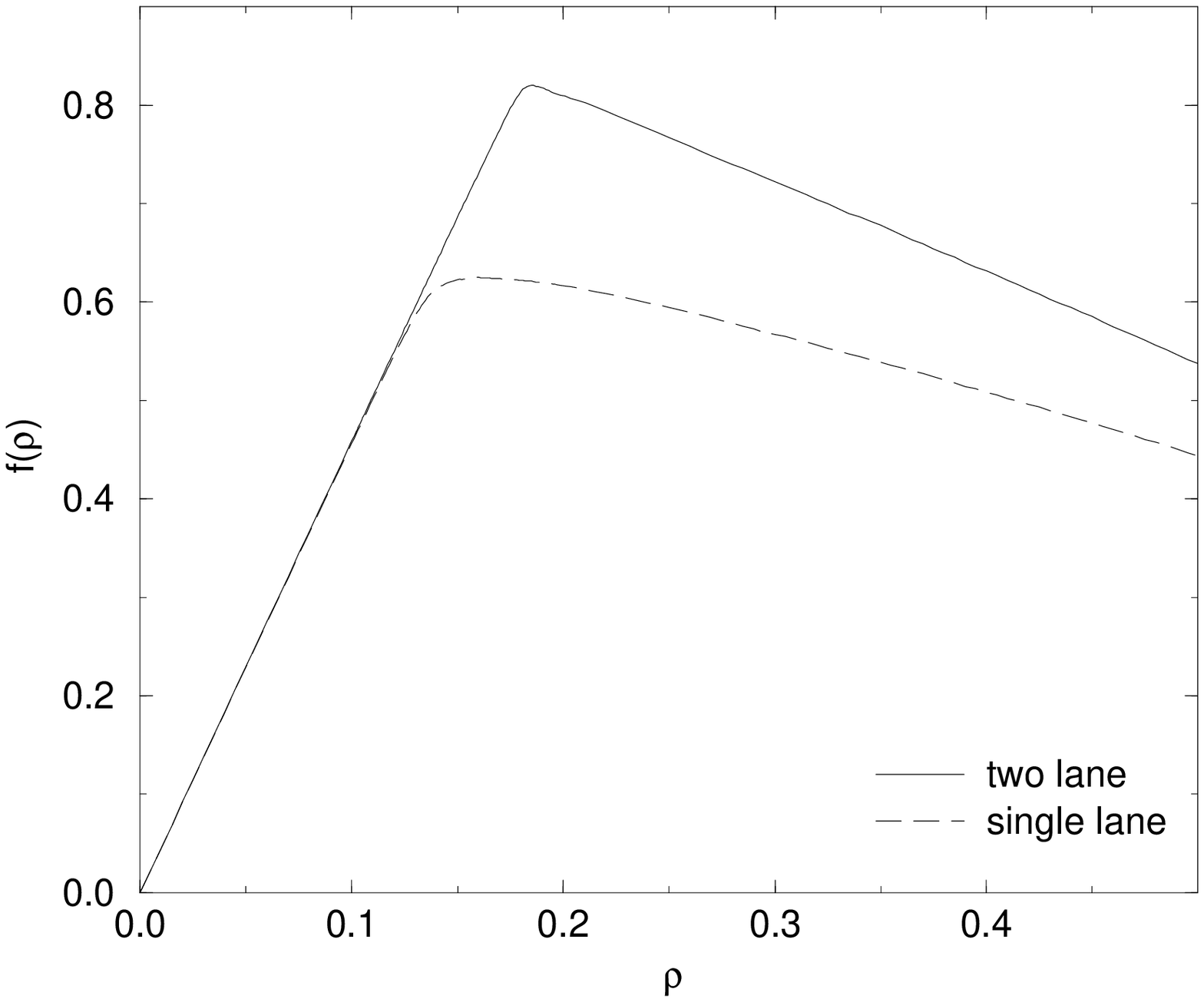}
\epsfxsize=\columnwidth\epsfbox{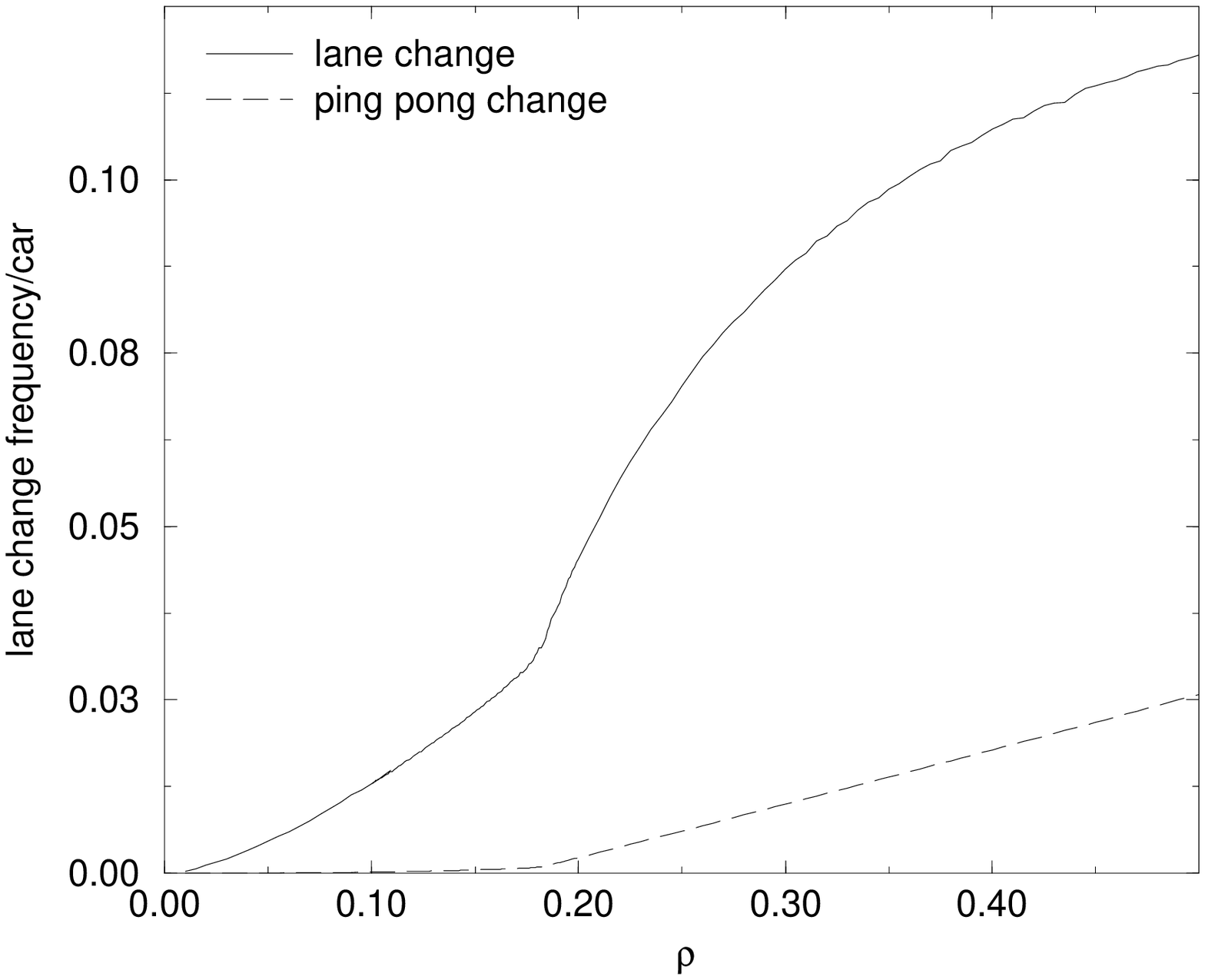}
\caption{Top: Fundamental diagram of the single-lane model
compared with the two-lane model for systems with $v_{max} = 5$ and $p= 0.4$.
Bottom: lane-changing and ping-pong change frequency in the two-lane model.}
\label{abbseq5}
\end{figure}

Compared to the NaSch model with parallel updating one obtains generically 
higher values of the average flow, if the same values of $v_{max}$ and 
$p$ are considered. Another feature of the sequential variant is the strong
$p$-dependence of $\rho_{max}$, e.g. in the limit $p \to 0$ the maximum
value of the flow is reached for $\rho \approx 1$. Nevertheless for
larger values of the braking noise one obtains quite realistic
fundamental diagrams as shown in Fig.~\ref{abbseq5}. 

After this short description of the single-lane behavior, we show
results for the two-lane system in the presence of particle
disorder. Obviously the lane-changing rules can be modified,
because the anticipation of the drivers allows for less restrictive safety
criteria. Therefore we introduced the following set of lane-changing
rules:
\begin{itemize}
\item Incentive criterion:
\begin{enumerate}
\item $\Delta v > gap$, \ \ \ with $\Delta v = v_{hope}-v_{same}$
\end{enumerate}
\item Safety criteria:
\begin{enumerate}
\item $gap_{other} \ge \Delta v$, \ \ \ with  $\Delta v = v - v_{other}$
\item $gap_{back} \ge \Delta v$, \ \ \ with  $\Delta v = v_{back} - v$
\end{enumerate}
\end{itemize}
Here $v_{same}$ denotes the velocity of the preceding car on the same lane.
$v_{other}$ and $v_{back}$ are the velocities of the neighbouring cars on
the destination lane.
$gap$, $gap_{other}$ and $gap_{back}$ have the same meaning as in the
definition given in Sec.\ \ref{Sec_Int}.

These lane-changing rules have the same structure as for the parallel
model. The first rule is the incentive criterion, which corresponds
to an unsatisfactory situation on the origin lane. The safety
criteria are necessary for a safe lane-changing maneuver. Obviously
the safety criteria differ drastically from those of the parallel case, 
e.g.\ if cars at the origin and destination lane move with the same velocity a
lane change is possible also for $gap_{other}=0$. These more aggressive
lane-changing rules lead to a much more efficient gap usage in the two-lane 
system. Compared to the single-lane system one obtains an
increase of the maximum flow of approximately $30\%$. Another effect of the
modified lane-changing rules is a strongly increased number
of lane changes, in particular the number of ping-pong lane changes
(i.e.\ lane changes of the same car in two consecutive time steps). 
Therefore the fraction of overtakings is reduced. Moreover it is 
also possible to change the lane at high densities.

\begin{figure}[hbt]
\epsfxsize=\columnwidth\epsfbox{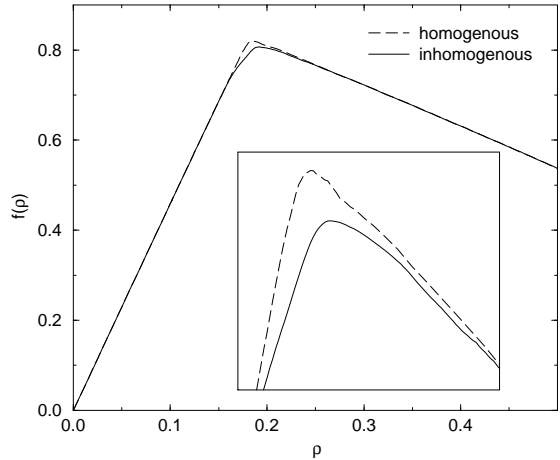}
\caption{Flow per lane of the homogeneous and the inhomogeneous
system with one slow car for the sequential model.}
\label{abbseq9}
\end{figure}

\begin{figure}[hbt]
\epsfxsize=\columnwidth\epsfbox{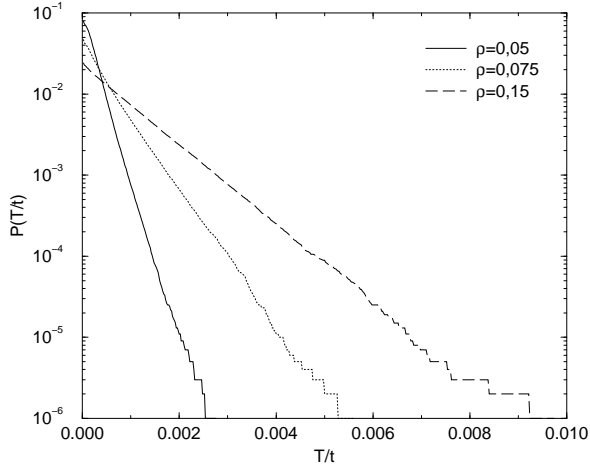}
\caption{Waiting time distribution in the sequential model with one slow
car.}
\label{abbseq9a}
\end{figure}

In contrast to the parallel model the flow only slightly changes in 
the presence of one slow car for densities near $\rho_{max}$ 
(Fig.~\ref{abbseq9}), while for all other densities the flow of the 
homogeneous system is recovered. Also the waiting time distribution 
(Fig.~\ref{abbseq9a})differs only slightly from the homogeneous
system, because lane-changing maneuvers can be performed more
effectively due to the less restrictive safety criteria.

\begin{figure}[hbt]
\epsfxsize=\columnwidth\epsfbox{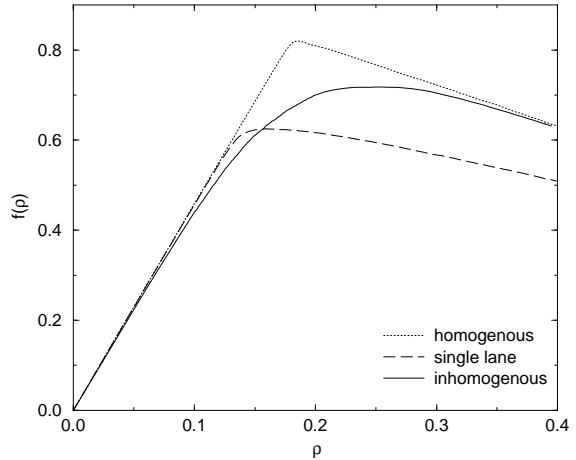}
\epsfxsize=\columnwidth\epsfbox{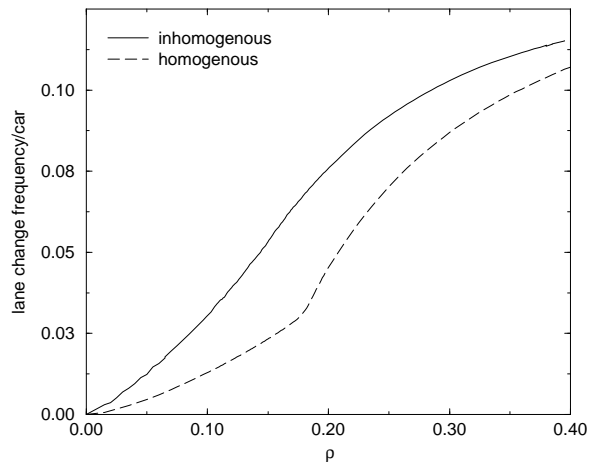}
\caption{Top: Comparison of the flow per lane of two-lane systems
with a single-lane system for the sequential update.
Bottom: Comparison of the lane change frequency of the inhomogeneous 
two-lane system with the homogeneous system ($p=0.4$).}
\label{abbseq14}
\end{figure}

Finally we investigated the case of $5\%$ slow vehicles in the system,
which was sufficient to dominate the system at low densities for the
case of parallel update. In contrast to the model introduced by
Rickert et al. \cite{Rickert2} only small deviations from the homogeneous 
case can be observed (Fig.~\ref{abbseq14}).

\subsection{Anticipation models}

Although the smoothing effect of the sequential update seems to be quite
satisfactory other features of this approach are clearly unrealistic.
Therefore we introduced a model variant with parallel update, where
anticipation effects are taken into account. Here smaller distances
between consecutive fast cars are made possible by estimating the
displacement of the predecessor in the next time step. The minimal
movement of the predecessor is given by $max(v_{next},0)$ where
$v_{next} = min(v_{pred}, gap_{pred})-1$ where $v_{pred}$ and $gap_{pred}$
denote the velocity and the gap in front of the predecessor.
This knowledge allows the introduction of an effective gap between the
cars which is given by $gap_{eff} = gap + v_{next}$. 

The anticipation leads to a slightly increased value of the maximum
flow, but major parts of the fundamental diagram are left unchanged. In
particular also for small values of the braking noise the fundamental
diagram is still quite close to the original model.

This modified version of the NaSch model allows for a set of lane-changing
rules analogous to the sequential case:

\begin{itemize}
\item Incentive criterion:
  \begin{enumerate}
    \item $(v > v_{same}$ and $gap < v)$ or $(gap_{other} > gap)$
  \end{enumerate}
\item Safety criteria:
\begin{enumerate}
\item[2.] $(v_{other} > v)$ or $(gap_{other} > gap)$
\item[3.] $v_{back} \le gap_{back}$
\end{enumerate}
\end{itemize}
All quantities have the same meaning as in the definitions given in previous
sections. Now lane changes are possible at small distances, e.g. only one
empty site between the leading car on the other lane and the changing
car is needed if both cars move with the same velocity. Again the
minimal gap is determined by velocity differences rather than by the
absolute value of the velocity. 

\begin{figure}[hbt]
\epsfxsize=\columnwidth\epsfbox{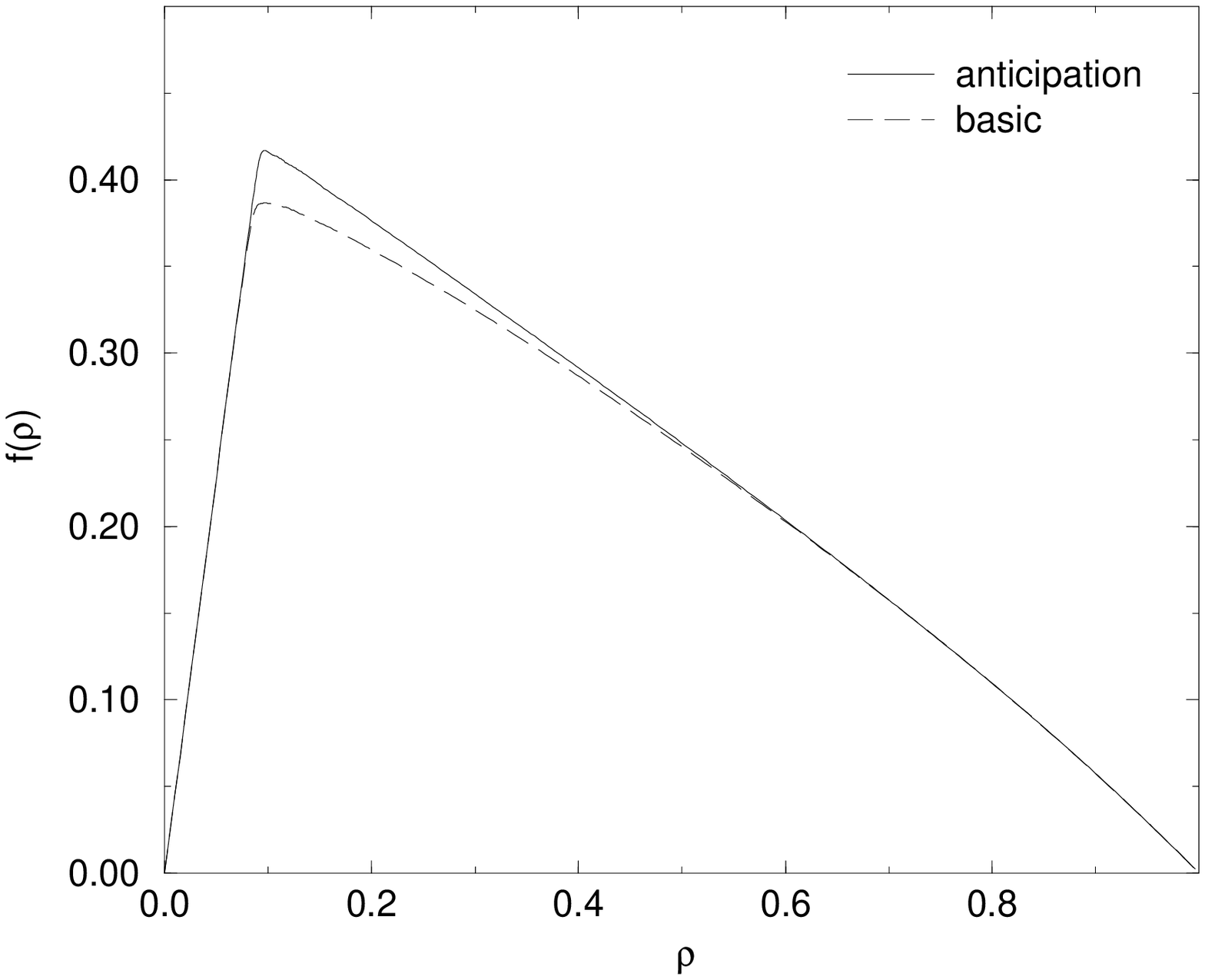}
\epsfxsize=\columnwidth\epsfbox{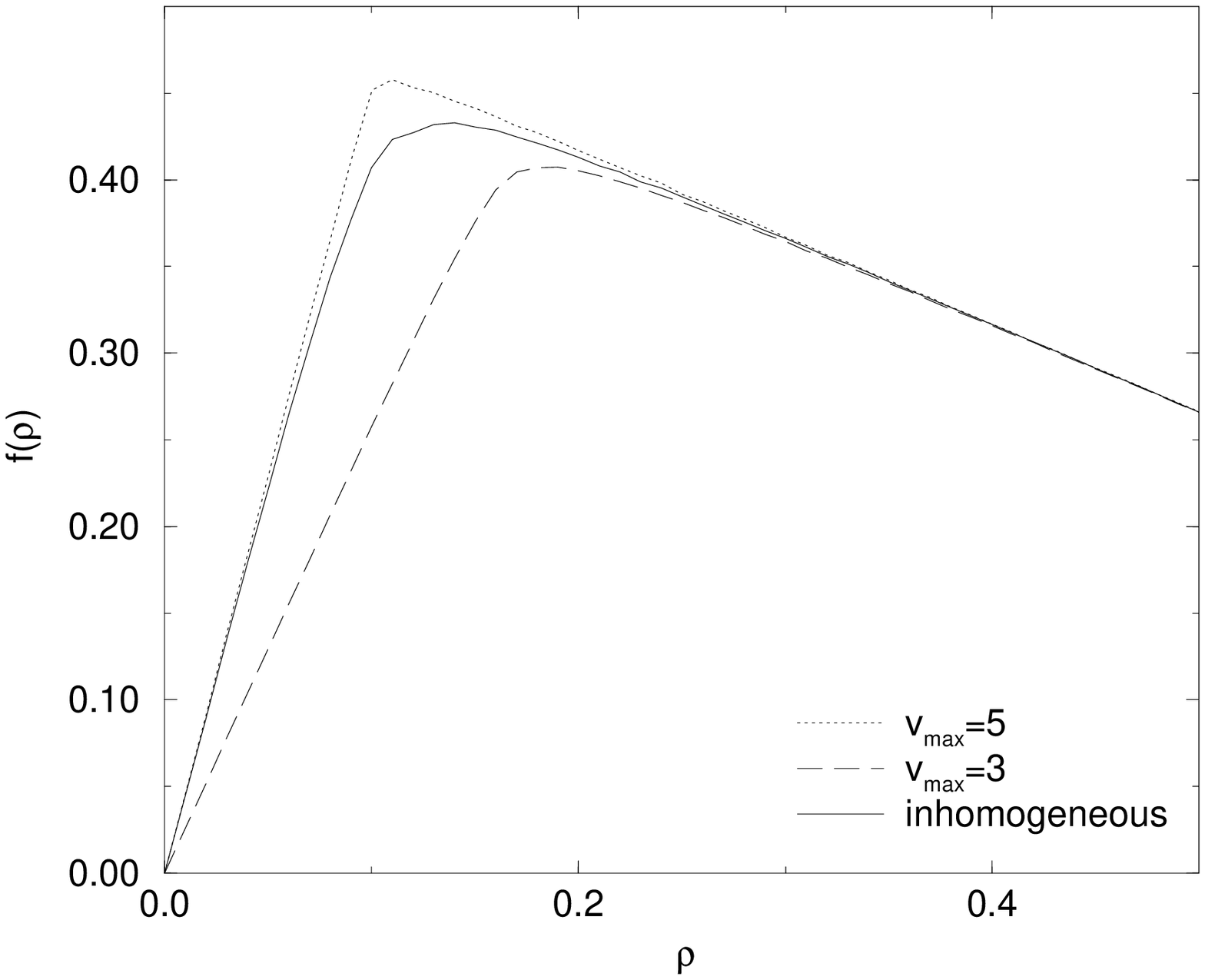}
\caption{Top: Comparison of the single-lane ``anticipation`` and the
  basic model.
 Bottom: Average flow of ``anticipation`` two-lane models ($p=0.4$).}
\label{anti}
\end{figure}

The simulation results show that such an implementation of anticipation 
effects is already sufficient to suppress the drastic influence
of slow cars (Fig.~\ref{anti}). Also if $5\%$ of slow cars are included, a
reduction of the flow is observable only close to $\rho_{max}$. The
increased robustness of the modified model shows that anticipation
effects are essential for a realistic description of multi-lane
traffic. 

\subsection{Other models}

We also investigated several model variants which interpolate
between the sequential and the anticipation models. We focussed
on models where the lanes are updated sequentially, but the update
within the lanes is parallel as in the NaSch model. This allows
to incorporate lane-changing rules which are more aggressive than
those of Rickert et al. \cite{Rickert2}.

In the simplest case the lane-sequential update is divided into four
substeps: a) lane-changes from right to left, b) update of the right lane,
c) lane-changes from left to right, d) update of the left lane.
These steps can be carried out in different order, e.g.\ a-b-c-d,
b-a-d-c, or a-b-d-c. The first two orderings have the disadvantage
that ping-pong lane-changes without forward motion or double updates
of cars are possible. The results of the simulations
show, however, that there are only small differences of the fundamental
diagram and the number of lane changes compared with the parallel update.
Furthermore the system is still very sensitive to a small fraction of
slow cars.

In all of these model variants cars move only `sidewards' during the
lane-changing substeps. If one allows cars also to move forward during 
lane-changes, the situation is improved slightly. Although the
cars still need the same safety distances for a lane-change, gaps can
be used more efficiently.

Finally, we introduced {\em temporary} anticipation into the basic model
of section \ref{Sec_models}.
Cars that changed the lane or the cars directly behind a vehicle that
changed to the other lane are allowed to anticipate for the next $n$
timesteps.
Therefore smaller safety distances are possible during lane changes.
This reduces the effects of disorder. In comparison to the anticipation
model of the previous subsection, the platoons forming behind slower
cars are larger and dissolve slower. If the cars stop anticipating after $n$
timesteps they usually have a short distance to the preceding car and are
likely to brake in the next update step due to the deceleration rule of the
NaSch model, i.e.\ temporary anticipation artificially causes jams.
In the limit $n\to\infty$ one recovers the fundamental diagrams of the 
anticipation model of the previous subsection.
In general, in order for anticipation to be most effective, all cars
have to anticipate.

\section{Summary and Discussion}

We have shown that even in two-lane systems, where fast
cars can overtake slow cars, particle disorder may dominate the
behaviour at low densities. 
The strongest influence of slow cars has been observed for
the NaSch model with symmetric lane-changing rules. For this model
variant already one slow car leads to platoon formation on both lanes,
i.e. the gap usage at low densities is not very effective. The platoon
formation has obviously drastic influence on the performance of the
system. If $5\%$ slow cars are included the performance of the
heterogeneous system is quite close to a homogeneous system of {\em slow} 
cars.  For asymmetric lane-changing rules slow cars influence
the systems performance less than in the symmetric case, because the
density on the ''fast'' (left) lane is reduced drastically, an
effective overtaking is possible. Therefore the flow reduction due to
the slow cars has a larger magnitude only close to the maximum.

Although these results look quite realistic, the distribution
of cars on both lanes differs drastically from those of real traffic.
Measurements on german highways show \cite{leutz,spar} that already for
densities below $\rho_{max}$ a higher density on the ''fast'' lane is
observable (lane inversion). 

Therefore one has to look for model variants which allow for lane
changes where the application of less restrictive safety criteria
is possible. An important feature
of this kind of models probably is the anticipation of the behavior
of the predecessor. Via calculating the velocity of the leading car,
one can accept much smaller distances than in the NaSch model. 

Anticipation is most effectively implemented with a (particle)
sequential update against the driving direction. The simulation
results show indeed a drastically reduced influence of slow cars, even
for symmetric lane-changing rules. Analogous results can be found for
a {\em parallel} model variant where the minimal movement of the
predecessor in the next time step is taken into account in order to
calculate an effective gap. Compared to the sequential model the
single-lane results for this anticipation model are much more realistic 
and the robustness of the two-lane system against particle disorder is
even larger than for the sequential variant.

We also studied the effects of a small fraction of fast cars in a system 
of slow vehicles. The fundamental diagram is almost identical to that 
of a pure system of slow cars. The average velocity of the fast cars 
is larger than that of the slow cars only for small densities.

In conclusion we have shown that particle disorder can lead to platoon
formation even in two-lane systems. Having in mind that also non-trivial 
lattice geometries may have strong effects on the macroscopic
behavior of the system, a very careful interpretation of traffic data
is necessary in order to distinguish between dynamical, multi-lane,
and boundary effects \cite{synchro}.\\[0.4cm]
\noindent{\bf Acknowledgments:}\\[0.1cm]
We like to thank Debashish Chowdhury for valuable comments on the manuscript.
The work of LS and AS has been performed within the research program of
the SFB 341 (K\"oln--Aachen--J\"ulich).


\end{document}